\documentclass[a4paper,12pt]{article}

\usepackage{graphicx}
\newcommand{\Slash}[1]{\ooalign{\hfil/\hfil\crcr$#1$}}
\def\wt{\widetilde}
\def\ov{\overline}
\def\simgt{\,{\rlap{\lower 3.5pt\hbox{$\mathchar\sim$}}\raise 1pt\hbox{$>$}}\,}
\def\simlt{\,{\rlap{\lower 3.5pt\hbox{$\mathchar\sim$}}\raise 1pt\hbox{$<$}}\,}

\def\gev{{\rm GeV}}
\def\tev{{\rm TeV}}

\def\ev{{\rm eV}}

\newcommand{\beq}{\begin{equation}}
\newcommand{\eeq}{\end{equation}}
\newcommand{\bea}{\begin{eqnarray}}
\newcommand{\eea}{\end{eqnarray}}
\newcommand{\bsub}{\begin{subequations}}
\newcommand{\esub}{\end{subequations} \noindent}
\newcommand{\clean}{\setcounter{equation}{0}}
\renewcommand{\theequation}{\thesection.\arabic{equation}}

\makeatletter
\newtoks\@stequation

\def\subequations{\refstepcounter{equation}%
  \edef\@savedequation{\the\c@equation}%
  \@stequation=\expandafter{\theequation}
  \edef\@savedtheequation{\the\@stequation}
  \edef\oldtheequation{\theequation}%
  \setcounter{equation}{0}%
  \def\theequation{\oldtheequation\alph{equation}}}

\def\endsubequations{%
  \ifnum\c@equation < 2 \@warning{Only \the\c@equation\space subequation
    used in equation \@savedequation}\fi
  \setcounter{equation}{\@savedequation}%
  \@stequation=\expandafter{\@savedtheequation}%
  \edef\theequation{\the\@stequation}%
  \global\@ignoretrue}

\def\eqnarray{\stepcounter{equation}\let\@currentlabel\theequation
\global\@eqnswtrue\m@th
\global\@eqcnt\z@\tabskip\@centering\let\\\@eqncr
$$\halign to\displaywidth\bgroup\@eqnsel\hskip\@centering
     $\displaystyle\tabskip\z@{##}$&\global\@eqcnt\@ne
      \hfil$\;{##}\;$\hfil
     &\global\@eqcnt\tw@ $\displaystyle\tabskip\z@{##}$\hfil
   \tabskip\@centering&\llap{##}\tabskip\z@\cr}

\makeatother

\begin{document}
\thispagestyle{empty}
\vspace*{-15mm}
\baselineskip 10pt
\begin{flushright}
\begin{tabular}{l}
{\bf OCHA-PP-184}\\
{\bf hep-ph/0112336}
\end{tabular}
\end{flushright}
\baselineskip 24pt 
\vglue 10mm 
\begin{center}
{\Large\bf	
	LEP2 constraint on 4D QED having dynamically generated 
        spatial dimension
}
\vspace{10mm}

\baselineskip 18pt 
\def\thefootnote{\fnsymbol{footnote}}
\setcounter{footnote}{0}
{\bf
Gi-Chol Cho$^{1)}$, Etsuko Izumi$^{2)}$ and 
Akio Sugamoto$^{1)}$}
\vspace{5mm}

$^{1)}$	{\it Department of Physics, Ochanomizu University, 
		Tokyo 112-8610, Japan}\\
$^{2)}${\it Graduate School of Humanities and Sciences,}  \\
{\it Ochanomizu University, Tokyo 112-8610, Japan}  \\
\vspace{2mm}

\vspace{10mm}
\end{center}
\begin{center}
{\bf Abstract}\\[7mm]
\begin{minipage}{14cm}
\baselineskip 16pt
\noindent
We study 4D QED in which one spatial dimension is dynamically generated 
from 3D action, following the mechanism proposed by Arkani-Hamed, Cohen 
and Georgi. 
In this model, the generated fourth dimension is discretized by an 
interval parameter $a$. 
We examine the phenomenological constraint on the parameter $a$ 
coming from the collider experiments on the QED process 
$e^+ e^- \to \gamma \gamma$. 
It is found that the LEP2 experiments give the constraint of 
$1/a \simgt 461\gev$. 
Expected bound on the same parameter $a$ at the future $e^+ e^-$ 
linear collider is briefly discussed. 
\end{minipage}
\end{center}

\vspace{5cm}
\begin{flushleft}
\end{flushleft}

\newpage
\baselineskip 18pt 
\section{Introduction}

The idea of dynamical generation of extra 
dimension~\cite{Arkani-Hamed:2001ca} has been applied to, 
for example, grand unified theories~\cite{deconstruction:gut}, 
electroweak symmetry breaking~\cite{deconstruction:ew}, and 
supersymmetry breaking~\cite{deconstruction:susy}, etc. 
From the high energy point of view, the extra dimension is 
generated dynamically at low-energy scale as a consequence of 
a certain fermion condensation mechanism~\cite{Arkani-Hamed:2001ca}. 
On the other hand, from the low-energy point of view, it seems 
that the space-time dimension will disappear above the ultraviolet 
cut-off scale, or, in other words, the asymptotic disappearance 
of space-time (ADST) occurs. 
In ref.~\cite{Sugamoto:2001uk}, the condensation mechanism is 
naively applied to the gauge theory in D dimensions, as well as 
the gravity theory in 3 dimensions. 
A characteristic feature of the generated (D+1) gauge theory 
or 4D gravity by this mechanism is that the generated 
dimension is discretized by an interval $a$\footnote{
The idea of the discretized extra dimension has been proposed 
independently by ref.~\cite{Hill:2000mu}.}. 
Application to gravity and the non-commutative geometry 
is further developed~\cite{Bander:2001qk, Alishahiha:2001nb}. 
   
In this paper, we study the effective 4D gauge theory in which 
one spatial dimension is dynamically generated from the 3D 
action~\cite{Sugamoto:2001uk}. 
We restrict our study to QED which is a simplest and well tested 
gauge theory. 
In general, due to the discretized spatial dimension, 
introduction of chiral fermions into the gauge theory causes 
so called ``doubling problem''~\cite{wilson,doubling problem}. 
Since there is no chiral fermion in QED, we are not troubled by 
this problem. 
This is another reason why we focus on QED in our 
study\footnote{
However, an approach to construct the effective 4D Standard 
Model with chiral fermions by controlling the doubling problem 
is proposed in ref.~\cite{Sugamoto:2001uk}. 
}. 
The appearance of the discretized extra dimension modifies 
the interactions between the photon and the electron. 
We first summarize the Feynman rules in the model and show that 
the modification occurs associated with the direction of 
the extra dimension and the interval parameter $a$. 
It is explicitly shown that the usual QED in 4D is restored 
in the continuum limit of the lattice in the extra dimension. 
It is, therefore, worth examining how large the size of $a$ is 
allowed phenomenologically.  
We study the annihilation process $e^+ e^- \to \gamma \gamma$ 
as a distinctive example of the QED process, and find 
quantitative constraint on the parameter $a$ from the measurement 
of the total cross section of the process at LEP2 experiments. 
The experimental data~\cite{Abbiendi:1999iu} tells us that 
the current constraint on the parameter $a$ is at most a few 
hundred GeV. 
We show that the future $e^+e^-$ linear collider may give 
a further bound on the parameter $a$ around TeV scale. 
   
In the next section, we briefly review the mechanism of dynamical 
generation of the extra dimension. 
The effective 4D QED generated from the 3D action based on the 
mechanism is also studied. 
In Sec.~3, the annihilation process $e^+ e^- \to \gamma \gamma$ 
is studied in the light of the measurement at LEP2. 
The last section is devoted to summary. 
\clean
\section{QED with the dynamical generation of extra dimension}
\label{intro}
\subsection{The dynamical generation of extra dimension}
\label{sub:one}
\begin{figure}[ht]
\begin{center}
\includegraphics[width=11cm,clip]{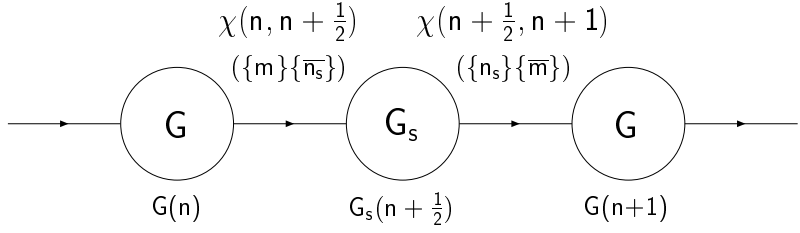}
\caption{
Diagrammatic description of the representation of the fermion 
$\chi$ under the two neighboring gauge groups, $G={\rm SU}(m)$ 
and $G_s={\rm SU}(n_s)$. 
The outgoing fermion $\chi(n,n+\frac{1}{2})$ from $G(n)$ is in 
the fundamental and the anti-fundamental representations of SU($m$) 
at the site $n$ and SU($n_s$) at the site $n+\frac{1}{2}$, 
respectively. 
On the other hand, the outgoing fermion $\chi(n+\frac{1}{2},n)$ 
from $G_s(n+\frac{1}{2})$ is in the fundamental and the 
anti-fundamental representations of SU($n_s$) at the site 
$n+\frac{1}{2}$ and SU($m$) at the site $n+1$, respectively. 
}
\label{fig:moose}
\end{center}
\end{figure}

We briefly review how 4D gauge theory is derived from the 3D 
action following refs.~\cite{Arkani-Hamed:2001ca, Sugamoto:2001uk}. 
First let us consider the following 3 dimensional gauge theory: 
It consists of $N$ sets of the gauge groups $G$'s labelled by 
$G(n), G(n+1), \cdots$, located at integer sites $n,n+1,\cdots$, 
and $G_s$'s labelled by $G_s(n+\frac{1}{2}), G(n+\frac{3}{2}), 
\cdots$, located at half-integer sites $n+1/2, n+3/2, \cdots$. 
The gauge coupling constants of $G$'s and $G_s$'s are 
$g$ and $g_s$, respectively. 
The Weyl fermion $\chi(m, \ell)$ belongs to a fundamental 
representation of $G_{(s)}(m)$ at the site $m$, and an 
anti-fundamental representation of $G_{(s)}(\ell)$ at the site 
$\ell$. 
The sites $m$ and $\ell$ always differ by $\frac{1}{2}$, so 
that the fermion $\chi$ connects the neighboring $G$ and $G_s$.  
In the following, we fix $G={\rm SU}(m)$ and $G_s={\rm SU}(n_s)$ 
for definiteness. 
We depict the diagrammatic description of the model in 
Fig.~\ref{fig:moose}. 
The outgoing fermion from $G(n)$, $\chi(n,n+\frac{1}{2})$, is 
in the fundamental representation of SU($m$) at the site $n$ 
and the anti-fundamental representation of SU($n_s$) at 
$n+\frac{1}{2}$. 
On the other hand, the outgoing fermion from $G_s(n+\frac{1}{2})$,  
$\chi(n+\frac{1}{2},n)$, is in the fundamental and the 
anti-fundamental representations of SU($n_s$) at $n+\frac{1}{2}$ 
and SU($m$) at $n+1$, respectively. 
The gauge coupling $g_s$ becomes strong below a certain low energy 
scale of $\Lambda_s$ and, as a result, it leads to the fermion 
condensations whose vacuum expectation values are given by the 
$m\times m$ unitary matrix $U$: 
\begin{eqnarray}
\frac{1}{2\pi (f_{s})^{2}} 
\langle \chi(n, n+\frac{1}{2})\chi(n+\frac{1}{2}, n+1)\rangle 
&\equiv& U(x; n, n+1), 
\label{eq:condensation}
\end{eqnarray}
where $f_s$ is the decay constant. 
Now $U(x; n, n+1)$ plays the role of a link field between 
$G(n)$ and $G(n+1)$. 
We identify the set of discrete points connected linearly 
by the link fields to the extra dimension. 
Then, the low-energy effective action below $\Lambda_s$ 
is given as follows
\begin{eqnarray}
S_{\rm eff} &=& 
-\frac{1}{2(g_3)^2} 
\int d^3 x ~~a\sum_{n=1}^{N}~~{\rm Tr}
(F_{ij}(x, n) F^{ij}(x, n)) 
\nonumber \\
&+& f_{s}\int d^{3}x ~\sum^{N}_{n=1} {\rm Tr}
\left[\left(D_{i}U(x; n, n+1)\right)^{\dagger}\left(D^{i}U(x; n,
n+1)\right)\right], 
\label{eq:effaction01}
\end{eqnarray}
where $i,j=0,1,2$. 
The first line of (\ref{eq:effaction01}) is the action of 
3D gauge field, while the second line is the action for the 
dynamically generated link field $U(x;n,n+1)$. 
The covariant derivative in (\ref{eq:effaction01}) is 
defined by 
\begin{eqnarray}
D_i &\equiv& \partial_i +iA_i(x, n)U(x; n, n+1)
 -iU(x; n, n+1)A_i(x, n+1). 
\end{eqnarray}
It is easy to see that eq.(\ref{eq:effaction01}) is 
essentially the 4D action for the gauge field of $G$, 
where the third spatial dimension is discretized. 
By noting $U(x; n, n+1)=e^{iaA_3(x:n)}$ and expanding 
eq.(\ref{eq:effaction01}) in a small dimensionful 
parameter $a$, the action could be written as 
\begin{eqnarray}
S_{\rm eff} &=& 
-\frac{1}{2(g_{3+1})^2} 
\int d^3 x ~~a\sum_{n=1}^{N}~~{\rm Tr}
(F_{\mu\nu}(x, n) F^{\mu\nu}(x, n)) 
\nonumber \\
&&~~ + \mbox{(higher order terms in $a$)}, 
\label{eq:effaction02}
\end{eqnarray}
where $\mu,\nu=0\sim 3$, and the 4D gauge coupling $g_{3+1}$ 
is given in terms of the 3D coupling $g_3$ as 
\begin{eqnarray}
(g_{3+1})^2 &=& a(g_3)^{2}, 
~~~~
a = \frac{1}{g_{3}(f_s)^{1/2}}. 
\end{eqnarray}
\subsection{Effective 4D QED}
\label{sub:two}
Let us write down the effective 4D QED having the dynamically 
generated extra spatial dimension following Sec.~\ref{sub:one}. 
The action for the gauge kinetic term, $S_g$, 
is already shown in the previous subsection 
for SU(n) gauge group (see, eq.~(\ref{eq:effaction02})). 
In the case of QED, $S_g$ is written as 
\begin{eqnarray}
S_g=-\frac{1}{4}  \int d^{3}x~~ a \sum_{n=1}^N 
F_{\mu\nu}(x,n) F^{\mu\nu}(x, n). 
\label{eq:qed}
\end{eqnarray}
   
The gauge invariant action $S_f$ for the fermion $\psi$ in 3D 
is given as; 
\begin{eqnarray}
S_f &=& \int d^3x \sum_{n=1}^N \Bigl(2aK\bar{\psi}(x, n)
i\gamma^{i}D_{i}(A(n))\psi(x, n) 
-\bar{\psi}(x, n)\psi(x, n) \Bigr), 
\label{action:fermion0}
\end{eqnarray}
where $i,j=0,1,2$ and we introduce the hopping parameter $K$ 
instead of the fermion mass parameter. 
The covariant derivative in (\ref{action:fermion0}) 
is defined by 
\begin{eqnarray}
D_i &\equiv& \partial_i+ieA_i. 
\end{eqnarray}
After the fermion condensation occurs as in 
eq.~(\ref{eq:condensation}), the link field $U(x;n,n+1)$ 
connects two fermions at the different sites $n$ and $n+1$. 
Then, the following pieces are also gauge invariant in addition 
to (\ref{action:fermion0}); 
\begin{subequations}
\begin{eqnarray}
\Delta_1 S_f &=& \int d^3x \sum_{n=1}^N K 
    \Bigl(\bar{\psi}(n)i\gamma^3 U(n, n+1)\psi(n+1) 
\nonumber \\
&&~~~~~~~
     -\bar{\psi}(n+1)i\gamma^3 U(n, n+1)^{\dagger} \psi(n) 
     \Bigr), 
\label{action:fermion1}
\\
\Delta_2 S_f &=& \int d^3x \sum_{n=1}^N K 
\Bigl( \bar{\psi}(n)U(n, n+1)\psi(n+1)
\nonumber \\
&&~~~~~~~
+\bar{\psi}(n+1) U(n, n+1)^{\dagger} \psi(n) \Bigr),  
\label{action:fermion2}
\end{eqnarray}
\label{action:fermion_both}
\end{subequations}
Here, we give a comment on the relative weight of these 
two terms, following Wilson in ref.~\cite{wilson}. 
If we choose the effective action as 
$\Delta_1 S_f +c \Delta_2 S_f$ with a relative weight $c$, 
then the dispersion relation of the electron becomes
\begin{equation} 
E^2-\left( \frac{\sin p_3 a}{a}\right)^2-
\left(\frac{1-2Kc\cos p_3 a}{2aK}\right)^2=0, 
\end{equation}
where we have set $p_1=p_2=0$ for simplicity.  
Using the variable $x$ defined by $x=\cos p_3 a$, the dispersion 
relation is generally the second order algebraic equation, 
namely, 
\begin{equation}
(c^2-1)x^2 - \left( \frac{c}{K} \right) x + 1 + 
\left( \frac{1}{2K} \right)^2 - ( aE )^2 = 0.
\end{equation}
Therefore, when $c^2 \ne 1$, we have generally two different 
solutions for the dispersion relation of electron for each 
spin degree of freedom. 
Then, we  have too many degrees of freedom for electron.  
To avoid this difficulty, we have to choose $c=1$ or $-1$ 
and make the dispersion relation to be the first order 
algebraic equation for $x$.  
We have, therefore, adopted $c=1$ in the above effective actions.  
This comment suggests that the condensation mechanism to generate 
the effective actions is so arranged that it may satisfy this 
condition.


Identifying the link filed $U$ as the third spatial component 
of the photon field 
\begin{eqnarray}
U(x; n, n+1)=e^{iaeA_3(x:n)}, 
\end{eqnarray}
we obtain the effective 4D QED action for small $a$, having 
a discretized spatial dimension with the interval $a$. 
\subsection{Feynman rules}

The Feynman rules of the effective QED with a discretized 
spatial dimension could be found from the actions (\ref{eq:qed}), 
(\ref{action:fermion0}) and (\ref{action:fermion_both}). 
Because of the discretized spatial dimension, the Lorentz 
invariance is manifestly violated for the finite size of $a$ 
so that the interaction between photon and electron becomes 
different from the usual QED. 
\begin{itemize}
\item Photon propagator\\
The photon propagator is given by (\ref{eq:qed}). 
Taking account of the discrete dimension, the derivative 
of the gauge field associated with the third spatial axis 
should be replaced by the corresponding different operator. 
Then, the propagator is given as 
\bsub
\begin{eqnarray}
D_{\mu\nu}(k) &=& 
\frac{ -i g_{\mu\nu} }{\displaystyle{\sum_{i=0,1,2} 
k_i k^i - \left(\frac{2}{a}\right)^2\sin^2\left(
\frac{k_3 a}{2}\right)}}
\label{eq:photon01}
\\ 
&\equiv&
\frac{ -i g_{\mu\nu} }{\wt{k}_\mu^* \wt{k}^\mu}, 
\label{eq:photon02}
\end{eqnarray}
\esub
where the second term of the denominator in (\ref{eq:photon01}) 
reflects the latticized dimension. 
The use of the notation 
\bea
\wt{k}_\mu &=& \left(\wt{k}_0,\wt{k}_1,\wt{k}_2,\wt{k}_3
\right)
\nonumber \\
&=&
\left(k_0,k_1,k_2,
\frac{2}{a} e^{-\frac{i}{2} k_3 a} 
\sin\left(\frac{k_3 a}{2}\right)
\right), 
\label{eq:photon_momentum}
\eea
allows us to have the familiar form of the photon propagator 
(\ref{eq:photon02}). 

\item Electron propagator\\
We find the electron propagator from 
eqs.~(\ref{action:fermion0}) and (\ref{action:fermion_both}), 
by rescaling the fermion fields canonically. 
The discrete dimension gives 
\bsub
\begin{eqnarray}
i S_F^{-1}(p) &=& 
\gamma^i p_i + \gamma^3 \frac{\sin\left(p_3 a \right)}{a}
- \frac{1}{2aK}\left\{(1-2K \cos(p_3 a) \right\}
\label{eq:electron01}
\\
&=& \Slash{\wt{p}} - m,  
\label{eq:electron02}
\end{eqnarray}
\label{eq:electron00}
\esub
where
\bsub
\begin{eqnarray}
\wt{p}_\mu &=& \left(\wt{p}_0, \wt{p}_1, \wt{p}_2, \wt{p}_3
\right)
\label{eq:electron_momentum}
\nonumber \\
&=&
\left(p_0,p_1,p_2,\frac{\sin(p_3 a)}{a} 
\right), 
\\
m &\equiv& \frac{1}{2aK}\left\{(1-2K \cos(p_3 a) \right\}. 
\end{eqnarray}
\esub
Note that the third component of the electron momentum 
$\wt{p}_3$ has a different form from the photon momentum 
$\wt{k}_3$ in eq.~(\ref{eq:photon_momentum}). 

\item electron-electron-photon vertex\\
The $ee\gamma$ vertex is modified associated with the 
non-canonical form in the third component of the momentum. 
The momentum dependent vertex function at the tree level 
is given by 
\bea
\Gamma^\mu &=& 
\left\{\gamma^i, \gamma^3\frac{1}{2}
\left( e^{-ip_3 a} + e^{+i (p_3+k_3) a} \right)
- \frac{i}{2}\left(e^{-i p_3 a } - e^{+i (p_3+k_3) a }
\right)\right\}, 
\label{eq:eegamma_vertex}
\eea
where $p$ and $k$ are the incoming electron and photon momenta, 
respectively. The outgoing electron momentum is $p+k$. 
\item Photon polarization sum and the spin sum of the spinors\\
It is rather straightforward to find the rules of the 
polarization sum of the photon and the spin sum of the spinors. 
They could be derived from the total action in the usual manner. 
The photon polarization sum follows 
\bea
\sum_{\lambda=1,2} \epsilon_{\mu}(\lambda) \epsilon_{\nu}^*(\lambda) 
&=& -g_{\mu\nu} - 
\frac{\wt{k}_\mu \wt{k}_\nu^*}
{\left| \wt{k}\cdot n \right|^2}
+ \frac{n_\mu \wt{k}_\nu^*}{\left(\wt{k}\cdot n \right)^*}
+ \frac{\left(n_\nu \wt{k}_\mu^* \right)^*}{\wt{k}\cdot n}, 
\label{eq:polarization}
\eea
where $n=(1,0,0,0)$. 
The spin sum of the spinors are given by 
\bsub
\bea
\sum_{{\rm spin}}
u(p)\ov{u}(p) &=& \Slash{\wt{p}} + m,  
\\
\sum_{{\rm spin}}
v(p)\ov{v}(p) &=& \Slash{\wt{p}} - m. 
\eea
\esub

\item Ward identity\\
The momentum dependence of the $ee\gamma$ vertex 
(\ref{eq:eegamma_vertex}) gives rise to a somewhat different 
form of the Ward identity. 
Taking account of the complex parameter $\wt{k}_\mu$ as 
the ``momentum'' of the photon, we find that there are 
two expressions of the Ward identity corresponding to 
the incoming and the outgoing photons: 
\bsub
\bea
-i \wt{k}_\mu \Gamma^\mu &=& S_F^{-1}(p+k)-S_F^{-1}(p), 
\label{ward01}
\\
+i \wt{k}^*_\mu \Gamma^\mu &=& S_F^{-1}(p-k)-S_F^{-1}(p), 
\label{ward02}
\eea
\label{ward00}
\esub
where (\ref{ward01}) and (\ref{ward02}) correspond to 
the vertices shown in Fig.~\ref{fig:ward}(a) and 
\ref{fig:ward}(b), respectively. 
Owing to these Ward identities, we understand that 
the effective 4D QED having dynamically generated spatial 
dimension is manifestly gauge invariant, even though it 
violates the Lorentz invariance. 
\begin{figure}[t]
\begin{center}
\includegraphics[width=10cm,clip]{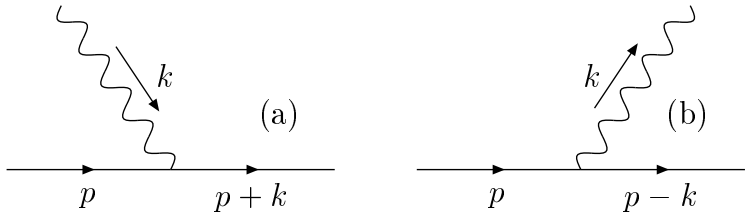}
\caption{The $ee\gamma$ vertices corresponding to the Ward 
identities in eq.~(\ref{ward01}) (a) and 
in eq.~(\ref{ward02}) (b). 
}
\label{fig:ward}
\end{center}
\end{figure}
\end{itemize}
\section{$e^+ e^- \to \gamma \gamma$ in the effective 4D QED}
\clean

In this section, we study the annihilation process 
$e^+ e^- \to \gamma \gamma$ taking account of the discrete 
third spatial dimension generated in the effective 4D QED. 
The axis $x^{3}$ representing this discrete dimension may be 
considered to be fixed in a special frame where the cosmic 
microwave background radiation is isotropic. 
Then, the beam axis of the $e^+ e^-$ collider is 
generally tilted from the $x^3$ axis by an angle $\chi$.  
Since the earth moves and rotates in this special frame for 
the cosmic background radiation, the angle $\chi$ changes 
in time. 
The main time variation of $\chi$ comes from the daily 
rotation of the earth (see \cite{Kamoshita} for the more 
detailed treatment on this problem). 
In this paper we estimate the total cross section averaged 
over the angle $\chi$ which can be compared with the ordinary 
total cross section measured by summing up the events obtained 
during the running time of the machine (several months or 
years). 
To clarify the day-night effect and the seasonal effect of 
the differential cross section of our process is an interesting 
problem to be pursued. 
Another possibility is that the discrete dimension is randomly 
directed at any place.  
Without referring to the details of the nature of the discrete 
dimension, that is, whether it is fixed in the special frame 
or is randomly directed, the averaging of the cross section 
over the tilt angle is a way to take into account the existence 
of the discrete dimension properly. 


As is shown in the previous section, the usual 4D QED 
is restored in the continuum limit of the third spatial 
dimension, or the $a \to 0$ limit. 
It is interesting, however, to examine how the cross section 
of the process is modified for the finite size of the interval 
parameter $a$, since $a$ is small but non-vanishing 
in the effective 4D QED. 
The presence of the finite $a$ leads to the explicit violation 
of the Lorentz invariance in 4D, but the evidence of the Lorentz 
symmetry violation has not been found yet. 
Therefore, we have to restrict our study to a small 
non-vanishing $a$. 
Furthermore, we take the massless electron limit for simplicity. 
In this limit, the gauge invariance requires the elimination of 
the part of the $ee\gamma$ vertex function (\ref{eq:eegamma_vertex}) 
not proportional to $\gamma_3$. 
Then, the Ward identity of the massless case is available in 
the original form (\ref{ward00}). 
For the incoming photon with the momentum $k$, the vertex 
function in the massless limit is given by 
\begin{eqnarray}
\Gamma^\mu &=& 
\left\{\gamma^i, \gamma^3\frac{1}{2}
\left( e^{-ip_3 a} + e^{+i (p_3+k_3) a} \right)
\right\}. 
\label{eq:vertex_massless}
\end{eqnarray}
   
\begin{figure}[t]
\begin{center}
\includegraphics[width=12cm,clip]{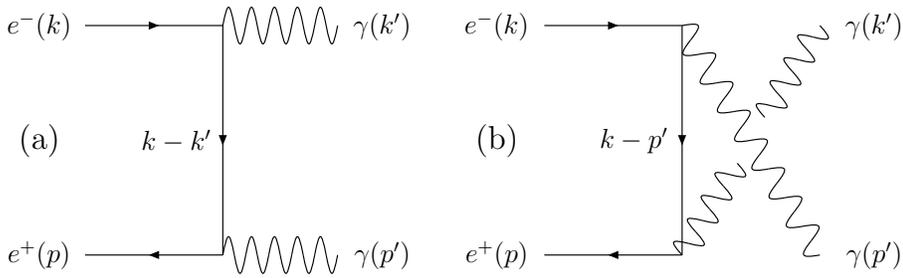}
\caption{Feynman diagrams for the annihilation process 
$e^+ e^- \to \gamma \gamma$. 
The $t$-channel and the $u$-channel diagrams are shown 
in (a) and (b), respectively. 
}
\end{center}
\label{fig:eegg}
\end{figure}

As is similar to the usual QED, there are $t$- and $u$-channel 
diagrams in the process $e^+ e^- \to \gamma \gamma$. 
The Feynman diagrams for both channels are shown in 
Fig.~\ref{fig:eegg}. 
The total amplitude is given by 
\bea
iM &=& \left(M^{\mu\nu}(t) + M^{\mu\nu}(u) \right)
\epsilon^*_\mu(p')\epsilon^*_\nu(k') , 
\eea
where $M^{\mu\nu}(t)$ and $M^{\mu\nu}(u)$ are the 
$t$- and $u$-channel amplitudes without the photon 
polarization vectors, respectively, and are found to be
\bsub
\bea
M^{\mu\nu}(t) &\equiv& 
    -e^2 \ov{v}(\wt{p}) \Gamma^\mu S_F(k-k') \Gamma^\nu u(\wt{k}), 
\\
M^{\mu\nu}(u) &\equiv& 
    -e^2 \ov{v}(\wt{p}) \Gamma^\nu S_F(k-p') \Gamma^\mu u(\wt{k}). 
\eea
\esub
From the gauge invariance, the following relations hold as in 
the usual QED when we employ $\wt{p}'$ and $\wt{k}'$ as the 
photon ``momentum''
\bea
\wt{p}'^*_\mu\left(M^{\mu\nu}(t)+M^{\mu\nu}(u)\right) 
&=& 
\wt{k}'^*_\nu\left(M^{\mu\nu}(t)+M^{\mu\nu}(u)\right) = 0. 
\eea
   
In practice, we take a small $a$ limit. 
Namely, we expand the photon momentum (\ref{eq:photon_momentum}), 
the electron momentum (\ref{eq:electron_momentum}) and 
the $ee\gamma$ vertex function (\ref{eq:vertex_massless}) 
in the parameter $a$, and consistently drop the higher order 
terms in the amplitudes. 

   
\begin{figure}[t]
\begin{center}
\includegraphics[width=11cm,clip]{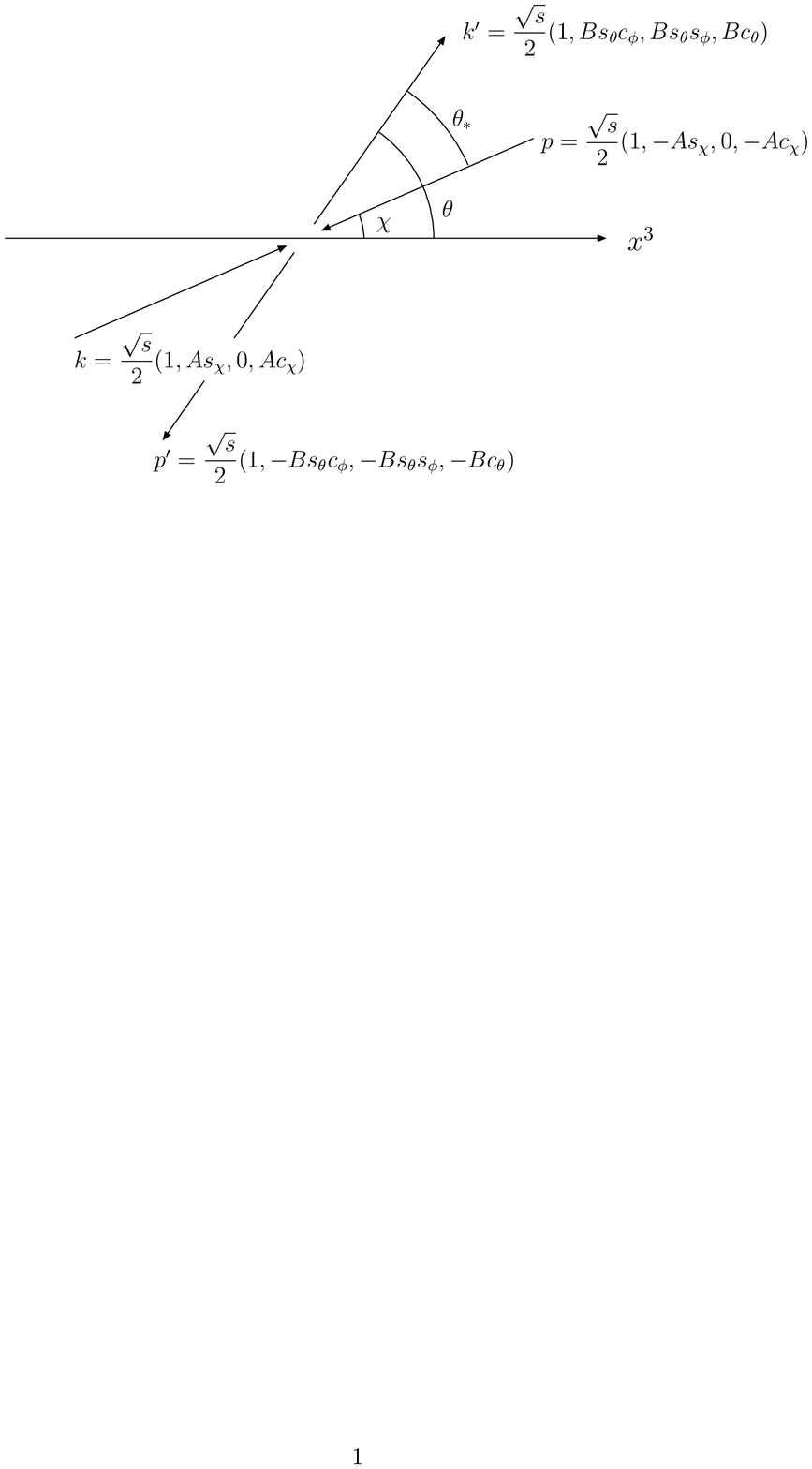}
\caption{Momentum assignment of the process 
$e^+ e^- \to \gamma \gamma$ in the CM system, where $x^{3}$ 
is the discrete dimension and the beam axis is tilted from 
the $x^{3}$ axis by the angle $\chi$.  
The scattering angle of the process is denoted by $\theta_{*}$. 
The parameters $A$ and $B$ account for the change of the 
dispersion relation in this model. 
The explicit forms of them are given in 
eq.~(\ref{eq:parameter.kinematics}). 
}
\end{center}
\label{fig:assignment}
\end{figure}
As is shown previously, the dispersion relations between 
the energy and the momentum for photon and electron take 
the different forms due to the finite interval parameter $a$. 
This tells us that the assignment of the four momenta in a 
specific frame is affected by the finite $a$-parameter. 
In Fig.~4, we show explicitly the momentum configuration of 
the relevant particles in the CM system. 
The four momenta receive the following corrections due to 
the finite $a$-parameter: 
\bsub
\bea
k &=& \frac{\sqrt{s}}{2}(1,As_\chi,0,Ac_\chi), \\
p &=& \frac{\sqrt{s}}{2}(1,-As_\chi,0,-Ac_\chi), \\
k' &=& \frac{\sqrt{s}}{2}(1,B s_\theta c_\phi,
       Bs_\theta c_\phi,B c_\theta), \\
p' &=& \frac{\sqrt{s}}{2}(1,-B s_\theta c_\phi,
       -Bs_\theta s_\phi,-B c_\theta), 
\eea
\label{eq:momentum:assign}
\esub
where $A$ and $B$ are defined as 
\bsub
\begin{eqnarray}
A(\chi) &\equiv& 1+ \frac{sa^2}{24}\cos^4\chi, 
\\
B(\theta)&\equiv&  1+\frac{1}{4}\frac{sa^2}{24} \cos^4\theta. 
\end{eqnarray}
\label{eq:parameter.kinematics}
\esub
Then, the squared amplitude is given by 
\bea 
\sum
|M|^2 &=& 8e^4 \left( 
                  \frac{1+\omega}{1-\omega}
                  +\frac{1-\omega}{1+\omega} 
                \right) 
\nonumber \\
          &+& 8e^4 \frac{sa^2}{12}
             \left [
                \frac{1}{(1-\omega)^2}
            (c_\chi-c_\theta)^4 + \frac{3}{4}(2c_\chi-c_\theta)^2
             \right.  
\nonumber 
\\ 
&&~~~~~ - \frac{1}{1-\omega}
            \left\{
            c_\chi^3 (c_\chi+c_\theta)-c_\theta(c_\chi-c_\theta)^3
+\frac{3}{2}(2c_\chi-c_\theta)^2(1-c_\chi c_\theta) 
            \right\}
\nonumber \\
&&~~~~~ +  \frac{1}{(1+\omega)^2}
            (c_\chi+c_\theta)^4 + \frac{3}{4}(2c_\chi+c_\theta)^{2} 
\nonumber \\
&&~~~~~\left. - \frac{1}{1+\omega}
            \left\{
c_\chi^3 (c_\chi-c_\theta)+c_\theta(c_\chi+c_\theta)^3
+\frac{3}{2}(2c_\chi+c_\theta)^2(1+c_\chi c_\theta) 
            \right\}
        \right.
\nonumber \\
&&~~~~~ \left. +\frac{1}{(1-\omega)(1+\omega)}3c_\theta^4 
        \right],
\label{eq:amp:squared}
\eea
where we summed over the spin of incoming electron and 
positron, and the polarization of outgoing photons.  
In the above expression, $\omega$ is defined as 
$\omega \equiv \cos\theta_{*}$ with the scattering angle 
$\theta_{*}$ of the process given in Fig.~4. 
The first line in r.h.s. is consistent with the usual 
QED, while the other lines correspond to the correction 
terms due to the latticized dimension. 
The second and third lines are found from the $t$-channel 
diagram, and the forth and fifth lines are from the 
$u$-channel diagram, while the last line represents the 
interference of the two diagrams, which must be absent 
in the usual QED. 
In the usual QED, it is well known that the squared amplitude 
of the $u$-channel diagram in the CM system is obtained from 
that of the $t$-channel diagram, by replacing the scattering 
angle $\omega=\cos\theta_{*}$ by $-\omega=-\cos\theta_{*}$ 
(see, the first line in (\ref{eq:amp:squared})). 
We find similarly that the correction terms of the $u$-channel 
(forth and fifth lines) are obtained from those of the 
$t$-channel (second and third lines) by the replacement 
$\theta \to \theta + \pi$, or $\chi \to \chi + \pi$.  
Either replacement induces $\omega \to -\omega$, since 
we have 
$\omega=c_{\theta_*}=c_\chi c_\theta + s_\chi s_\theta c_\phi$.
 The finite $a$ corrections appear also in the beam 
flux factor and in the phase space integration in the CM system.  
The beam flux $f$ depends on the relative velocity of the 
electron and the positron, namely 
\begin{equation}
f=sv_{\rm rel}=2s|v_{\rm electron}|.
\end{equation} 
The velocity $\mbox{\boldmath{$v$}}$ defined by 
$\mbox{\boldmath{$v$}}=\partial E/\partial 
\mbox{\boldmath{$p$}}$, depends on the modified dispersion 
relation of electron or positron, and so we have 
\begin{equation}
\frac{1}{f}=\frac{1}{2s}A^3,
\label{flux}
\end{equation}
which is valid in the lowest order corrections in $a$.  
The phase space integration $d\Phi$ is also modified as 
\begin{equation}
d\Phi = \frac{1}{16\pi} B^{5} d(\cos\theta)\frac{d\phi}{2\pi},
\label{ps}
\end{equation}
in the same lowest order approximation.


Now, the differential cross section of the process 
$e^+e^- \to \gamma \gamma$ in the CM system are given by
\begin{equation}
d\sigma(e^+e^- \to \gamma \gamma) = 
\frac{1}{64\pi s}A(\chi)^3 B(\theta)^5d(\cos\theta)\frac{d\phi}{2\pi} 
\frac{1}{4}\sum |M|^2 .
\end{equation}
The factor $\frac{1}{64\pi s}$ in the above formula is 
the product of $\frac{1}{2s}$ from eq.~(\ref{flux}), 
$\frac{1}{16\pi}$ from eq.~(\ref{ps}), and the Bose factor 
$\frac{1}{2}$ coming from the final two photons.
Here, we will take the average of the differential cross 
section over the angle $\chi$.  
For this purpose, it is convenient to move to another 
coordinate frame in which the beam axis of the incident electron 
points to the direction of the new $x^{3}_{*}$ axis and the 
scattering angle $\theta_{*}$ and $\phi_{*}$ are the polar angle 
and the azimuthal angle of the new coordinate frame.  
Then, the $\chi$ dependence disappears in the denominators made 
of $1 \pm \omega=1 \pm c_{\theta_{*}}$, and the $\chi$ dependence 
appears only in the numerators.  
Now, the average over $\chi$ is easily performed, by using 
the transformation rule 
$c_\theta = c_\chi c_{\theta_*} - s_\chi s_{\theta_*} c_{\phi_*}$ 
and the following formulas of the average:
\bea
\langle c_{\chi}^4 \rangle = \langle s_{\chi}^4 \rangle =\frac{3}{8}, 
~
\langle c_{\chi}^2 s_{\chi}^2 \rangle = \frac{1}{8}, 
~
\langle c_{\chi}^2 \rangle =\langle s_{\chi}^2 \rangle=\frac{1}{2}.
\eea
After taking the average over $\chi$ is finished, the integration 
over the azimuthal angle $\phi_*$ is easily performed.


Then, the total cross section averaged over the tilt angle 
$\chi$ of the process $e^+e^- \to \gamma \gamma$ in the CM 
system are obtained as (note $\omega \equiv \cos\theta_*$)
\begin{eqnarray}
\sigma 
&=&\frac{\pi \alpha^2}{s} \int d\omega
\left\{ 
   \frac{1+\omega^2}{1-\omega^2}
\right.  \nonumber 
\\
&+& \frac{sa^2}{24} \times 
\left[
   \frac{3}{16}\left(
       3+\frac{5}{32}(3+2\omega^2
      +3\omega^4 )
                \right) 
\times \left( 
  \frac{1+\omega}{1-\omega} 
+ \frac{1-\omega}{1+\omega}
       \right) 
\right. \nonumber 
\\
&+& \frac{3}{32}(55+7\omega^2)\nonumber 
\\
&+& \frac{1}{1-\omega}\frac{1}{64} 
    (237-438\omega + 210\omega^2
    -54\omega^3 + 9\omega^4 ) \nonumber 
\\
&+& \frac{1}{1+\omega}\frac{1}{64} 
    (237+438\omega + 210\omega^2
    +54\omega^3 + 9 \omega^4) \nonumber 
\\
&+& \left. 
         \left. 
          \frac{9}{64} \frac{8\omega^2
         +3(1-\omega^2)^2}{1-\omega^2} 
         \right] 
     \right\}.
\end{eqnarray}
\begin{figure}[t]
\begin{center}
\includegraphics[width=11cm,clip]{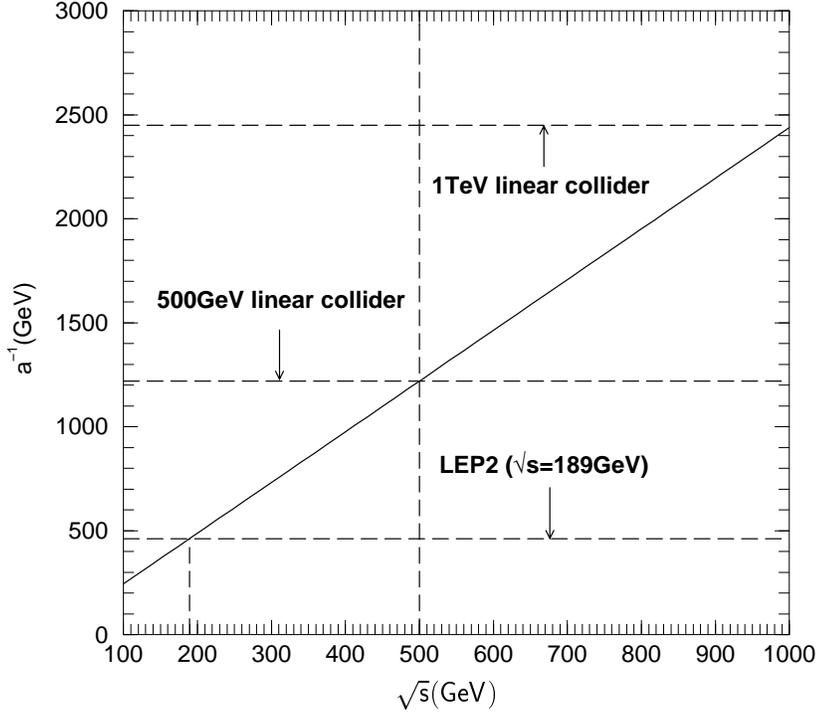}
\caption{The upper bound of the interval parameter $a$ 
as a function of $\sqrt{s}$, obtained from the 
total cross section of the process $e^+ e^- \to \gamma \gamma$.  
We assume that the cross section normalized by the QED prediction 
can be measured in the 5\% accuracy. 
The bottom horizontal line denotes the constraint on 
$a^{-1}$ from the LEP2 experiments. 
The middle horizontal line is the expected lower bound 
on $a^{-1}$ at $500\gev$ $e^+ e^-$ linear collider, while 
the top horizontal line corresponds to the bound on $a^{-1}$ at 
$1\tev$ $e^+ e^-$ linear collider. 
}
\label{fig:constraint}
\end{center}
\end{figure}
   
The total cross section of $e^+ e^- \to \gamma \gamma$ has been 
measured at LEP2~\cite{Abbiendi:1999iu}, in which 
the experiment has been done for the scattering angle 
$|\cos\theta_*|<0.9$. 
We find the total cross section within this range of 
$\cos\theta_*$ as 
\bea
\frac{\sigma}{\sigma_{\rm QED}} &\approx& 1+0.30 sa^2, 
\eea
where the cross section in QED is given by 
\bea
\sigma_{\rm QED} &=& \frac{\pi \alpha^2}{s}
\int d\omega \frac{1+\omega^2}{1-\omega^2}
\nonumber \\
&=&
4.09 \frac{\pi \alpha^2}{s}.
\eea
In ref.~\cite{Abbiendi:1999iu}, the total cross section is 
measured at $\sqrt{s}=189\gev$. 
When the measured total cross section is normalized by the QED 
prediction, the error is about $5\%$~\cite{Abbiendi:1999iu}. 
Then it constrains the parameter $a$ as 
\bea
0.30 sa^2 &\simlt & 0.05 ~~~\mbox{\rm for $\sqrt{s}=189\gev$}
\nonumber \\
&\Longrightarrow&  1/a \simgt 461\gev. 
\label{eq:const}
\eea
From its dimensionality, the parameter $a$ always appears 
as a product with the CM energy ($\sqrt{s}$) and the correction 
to the physical observables appears in ${\cal O}(sa^2)$. 
This is because $a \to -a$ means the parity transformation 
$p_3 \to -p_3$ which is kept invariant in this effective 4D QED. 
Then, the constraint on the $a$-parameter will be much stronger 
than (\ref{eq:const}) as the beam energy increases if the cross 
section is measured as precisely as the LEP2 
experiment~\cite{Abbiendi:1999iu}. 
We show in Fig.~\ref{fig:constraint} the parameter $a^{-1}$ in 
the unit of GeV as a function of the beam energy $\sqrt{s}(\gev)$. 
In the figure, we assume that the experimental error of the 
total cross section normalized by the QED prediction is $5\%$. 
The bottom horizontal line in the figure corresponds to the lower 
bound on $a^{-1}$ from the LEP2 experiment performed at 
$\sqrt{s}=189\gev$, eq.(\ref{eq:const}). 
As is mentioned, the lower bound on $a^{-1}$ significantly 
increases for the higher $\sqrt{s}$. 
We show the middle and the top horizontal lines as the expected 
constraints on $a^{-1}$ from the future $e^+ e^-$ linear collider 
at $\sqrt{s}=500\gev$ and $1\tev$, respectively. 
The constraint on the parameter $a$ is found to be
$1/a \simgt 1.2\tev$ at the $500\gev$ linear collider, 
while the $1\tev$ linear collider will cover the $a$ parameter 
less than $(2.4\tev)^{-1}$. 

\section{Summary}

We have studied the 4D QED in which the third spatial 
dimension is dynamically generated owing to the mechanism of 
the certain fermion-pair condensation proposed in 
ref.~\cite{Arkani-Hamed:2001ca}. 
The direct consequence of employing this mechanism is that 
the dynamically generated extra dimension is discretized by 
the interval parameter $a$. 
We have derived the effective 4D QED action explicitly, and 
shown the Feynman rules. 
They take different forms from those in the usual 
QED due to the latticized fourth dimension. 
Taking account of the discrete third spatial dimension, we 
examined the QED process $e^+ e^- \to \gamma \gamma$ and found 
the quantitative constraint on the interval parameter $a$ from 
the LEP2 experiment. 
In our estimation of the cross section of $e^+ e^- \to 
\gamma \gamma$, the beam axis of $e^{+}$ and $e^{-}$ is 
considered to be tilted from the discrete third spatial 
dimension, and the cross section is averaged over the 
tilt angle. 
This averaging is a way to take into account the existence of 
the discrete dimension properly, without referring the details 
of the nature of the discrete dimension, that is,  whether 
the discrete dimension is fixed in the special frame for 
the microwave background radiation, or the direction of it is 
randomly generated.  
Then, the constraint on the interval parameter is found 
to be $1/a \simgt 461\gev$, and the constraint will be stronger 
at future $e^+ e^-$ linear collider. 
If the linear collider could measure the total cross section 
of the process as accurately as in LEP2, the size of the 
lattice in the third spatial dimension will be bound as: 
$1/a \simgt 1.2\tev$ for $\sqrt{s}=500\gev$ and 
$1/a \simgt 2.4\tev$ for $\sqrt{s}=1\tev$. 
   
The finite size of $a$ modifies the dispersion relation, 
which indicates the violation of the Lorentz invariance. 
The possibility of violating the Lorentz symmetry has been 
examined, and two phenomena difficult to explain in modern 
astrophysics, 
such as the ultra high-energy cosmic rays beyond the GZK 
cut-off of $10^{20} \ev$, and $20\tev$ $\gamma$-ray from 
the active galaxy of Markarian 501, 
may be understood by the Lorentz symmetry 
violation~\cite{lorentz.violation}. 
The interpretation of these problems in the framework of our 
model will be done elsewhere~\cite{CIS}. 
\clean
\section*{Acknowledgements}

This work is supported in part by the Grant-in-Aid for Science 
Research, Ministry of Education, Science and Culture, Japan 
(No.13740149 for G.C.C. and No.11640262 for A.S.).


\end{document}